\documentclass[11pt]{article}

\usepackage[margin=1in]{geometry}
\usepackage{amsmath,amssymb,mathtools}
\usepackage{booktabs,array,longtable}
\usepackage{graphicx}
\usepackage{enumitem}
\usepackage{hyperref}
\usepackage{xcolor}
\usepackage{microtype}
\usepackage{float}   

\graphicspath{{figures/}}
\newcommand{\FloatBarrier}{}

\hypersetup{colorlinks=true,linkcolor=blue,citecolor=blue,urlcolor=blue,hypertexnames=false}
\setlist[itemize]{leftmargin=1.5em,itemsep=0.25em,topsep=0.25em}
\setlist[enumerate]{leftmargin=1.5em,itemsep=0.25em,topsep=0.25em}

\newcommand{\E}{\mathbb{E}}
\newcommand{\Cov}{\operatorname{Cov}}
\newcommand{\Var}{\operatorname{Var}}
\newcommand{\KL}{D_{\mathrm{KL}}}
\newcommand{\Stot}{S_{\mathrm{tot}}}
\newcommand{\Sdir}{S_{\mathrm{dir}}}
\newcommand{\Strue}{S_{\mathrm{true}}}
\newcommand{\Ssat}{S_{\mathrm{sat}}}
\newcommand{\Idir}{I_{\mathrm{dir}}}
\newcommand{\Itrue}{I_{\mathrm{true}}}
\newcommand{\phidir}{\phi_{\mathrm{dir}}}
\newcommand{\sigmoid}{\sigma}
\newcommand{\logit}{\operatorname{logit}}
\newcommand{\Pemp}{P_{\mathrm{emp}}}
\newcommand{\qemp}{q_{\mathrm{emp}}}
\newcommand{\qprod}{q_{\mathrm{prod}}}
\newcommand{\qbal}{q_{\mathrm{bal}}}

\title{Feature leakage and the identifiability of direct-dependency entropy models of neural activity}
\author{
Houman Safaai$^{1}$\thanks{Correspondence: \href{mailto:houman_safaai@harvard.edu}{houman\_safaai@harvard.edu}, \href{mailto:bernardo_sabatini@hms.harvard.edu}{bernardo\_sabatini@hms.harvard.edu}} \quad Bernardo L.\ Sabatini$^{1,2}$\\[0.5em]
{\small $^{1}$Kempner Institute for the Study of Natural and Artificial Intelligence at Harvard University}\\
{\small $^{2}$Department of Neurobiology, Howard Hughes Medical Institute, Harvard Medical School}
}
\date{}

\begin{document}
\maketitle

\begin{abstract}
Biological neurons receive thousands of synaptic inputs on branching, electrically excitable dendrites. Many local nonlinear interactions between coactive synapses have been described, yet population activity is often modeled with simpler direct input-output rules in which each input contributes independently to a scalar drive. A natural question is therefore not only whether such direct models predict neural activity, but what their success reveals about the computation carried out by the neuron. We study this question for conditional maximum-entropy (MaxEnt) models that match the output rate and pairwise output-input coactivities. These models provide a principled measure of prediction by first-order sufficient statistics. However, as we describe, the associated entropy calculation cannot be used as a mechanism-identification test. A restricted MaxEnt fit is an information projection under the sampled input distribution. Omitted interaction, temporal, or hidden-state terms can be absorbed into fitted first-order parameters whenever they are correlated with the included statistics, yielding models with high goodness of fit that are mechanistically incorrect. For sparse correlated binary inputs, this absorption has a simple coskewness form that we derive explicitly. We introduce diagnostics that separate in-distribution prediction from recovery of the response rule. These include (1) a state-reweighting control that holds \(P(y\mid x)\) fixed while changing \(P(x)\) (using a mild product-marginal control and a balanced stress-test endpoint); (2) conditional log-odds contrasts that test additivity of the local response table; and (3) a temporal leakage simulation that quantifies when instantaneous statistics absorb lagged mechanisms. In ground-truth simulations, purely higher-order responses can pass first-order entropy and raw coactivity tests under leakage-prone sampling, whereas the same response is correctly classified after reweighting. We apply the diagnostics to selected, leakage-enriched local tables from CA1 hippocampal recordings. We examine diagnostic subsets chosen to maximize first-order predictability, not as a representative survey of CA1 activity. In these tables, approximately half of those that appear first-order under empirical weights become distribution-sensitive under balanced reweighting, an effect far above a matched additive-surrogate null. An unselected control, analyzed conditional on empirical-success tables, shows that removing the output-based input selection does not remove the effect. This shows how large the absorption can be under the sampling geometry of the data to which these models are applied. In short, these models do not provide an estimate of the fraction of neural responses that are mechanistically higher-order. Direct entropy-explained fractions and raw coactivity predictions are therefore best interpreted as predictions under the observed state distribution, not as evidence that mechanisms outside the direct model are absent or small.
\end{abstract}

\section{Introduction}

Biological neurons integrate synaptic inputs on complex dendritic trees, where voltage-dependent conductances, local synaptic interactions, plateau potentials, and spatially-confined integration can make the response to coactive inputs nonlinear \cite{Poirazi2003,Polsky2004,LondonHausser2005,Smith2013}. At the same time, much of theoretical neuroscience and machine learning uses a simpler description of neural computation: each input contributes independently to a summed drive, and the neuron applies a scalar nonlinearity to produce an output. How much of the activity of an individual neuron results from interactions between its inputs is unknown. Answering this question will require obtaining large-scale simultaneous recordings of the activity patterns of an individual neuron and all of its inputs. The need to model such data motivates the statistical and computational question: if an additive input-output model (i.e. a linear model with no interactions between inputs) predicts neuronal activity well, what, if anything, has been learned about the underlying mechanism that generated the neural response?

Conditional maximum entropy offers a potential method to address the predictive part of this question. In these MaxEnt approaches, one models a binary output by the least-structured conditional distribution that matches the output rate and the pairwise output-input coactivities, and examines how much activity remains unexplained. A recent application of MaxEnt to hippocampal, visual cortex, and \emph{C. elegans} neural activity reported that such direct models predict much of the raw higher-order and time-delayed coactivities in the datasets \cite{Lynn2026}. Those results are important as statements about prediction. The harder question is whether they identify the absence, or relatively small contributions, of mechanisms outside the direct model.

The gap between prediction and interpretation is familiar from the study of model misspecification \cite{White1982,Huber1967}: a restricted model can fit data well for the wrong structural reason. The issue is sharpened by the empirical distribution of recorded input states. Entropy explained by a direct model, and prediction of raw coactivities, are averages over that distribution. If omitted interaction or temporal features are correlated with the direct sufficient statistics on the sampled states, their effects can be folded into the fitted first-order parameters.

This problem is especially acute for sparse binarized recordings. Binarization collapses positive powers of a single input because \(x^n=x\) on \(\{0,1\}\) for all \(n\geq1\), and correlations among sparse binary inputs create a coskewness channel through which interaction terms project onto first-order terms. Sparse high-dimensional data can also downweight or miss the input states that distinguish an additive response from an interaction response. As a result, the same conditional response can yield very different first-order attribution under different state weights, even though the response mechanism itself has not changed.

Here, we develop an identifiability framework for direct-dependency conditional MaxEnt models of neural activity. We first use an information-projection calculation to show how omitted statistics can shift fitted first-order MaxEnt parameters and leave only residual structure detectable (Eq.~\ref{eq:fisher_leakage}). We then derive a coskewness identity for sparse binary inputs, which makes the projection mechanism explicit (Eq.~\ref{eq:binary_coskew}). These calculations motivate three diagnostics: state reweighting, conditional log-odds contrasts, and a temporal leakage control. We apply the diagnostics to ground-truth simulations and to publicly available CA1 and visual-cortex recordings.

The framework is not meant to dismiss first-order models, and it certainly does not imply that higher-order mechanisms dominate every dataset. It provides quantitative tools for stating the predictive contribution of a direct-dependency MaxEnt model with appropriate scope. In summary, prediction by first-order sufficient statistics under \(\Pemp(x)\) is a well-defined and useful descriptor, but it does not identify the absence of mechanisms outside the direct model. It cannot distinguish, on its own, whether biological neurons nonlinearly integrate synaptic information or are effectively additive logistic units under the sampled conditions. Related cautions about restricted maximum-entropy population models, coupled GLMs, higher-order correlations, dendritic nonlinearities, and activity-based connectivity inference appear in prior work \cite{Schneidman2006,Pillow2008,Roudi2009,Fitzgerald2011,Montani2009,DasFiete2020,Polsky2004,LondonHausser2005,Poirazi2003}; the framework here addresses identifiability in a unified way and applies to any direct-dependency MaxEnt analysis of binarized neural data.

\begin{figure}[tbp]
  \centering
  \includegraphics[width=\textwidth]{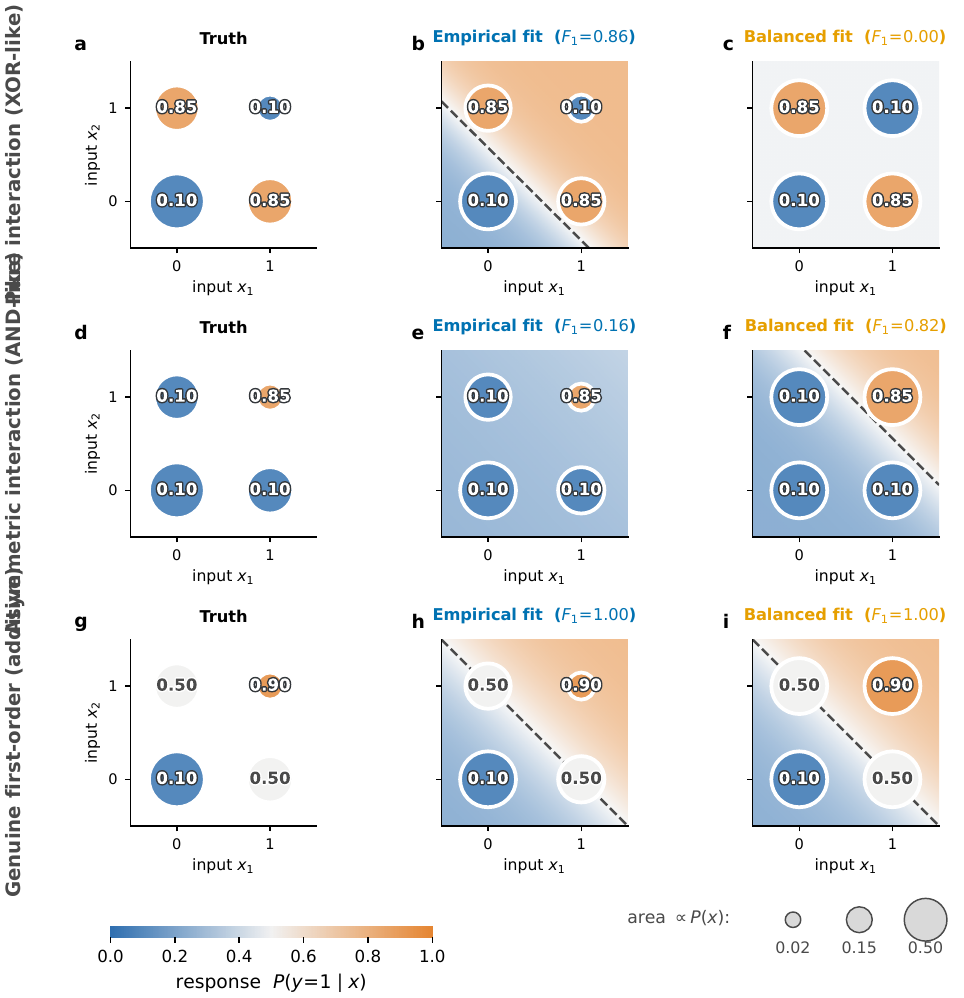}
  \caption{
  State weighting can determine whether a fixed response table appears first-order. Bubble color is the true response \(P(y=1\mid x)\), bubble area is the state weight \(P(x)\), shaded backgrounds are best first-order readouts and dashed lines mark the \(\hat P=0.5\) boundary. \textbf{a--c} A pure XOR-like interaction is well described by a first-order readout under empirical rare-corner sampling (\(F_1=0.86\)) but not under balanced sampling (\(F_1=0.00\)). \textbf{d--f} An asymmetric AND-like interaction shows the opposite masking pattern: empirical sampling misses the high-response corner, while balanced sampling recovers a strong first-order readout. \textbf{g--i} A genuinely additive response remains first-order under both samplings. Thus the diagnostic separates stable first-order response form from first-order attribution that depends on the sampled state distribution.
  }
  \label{fig:concept}
\end{figure}
\FloatBarrier

\section{Background: direct conditional MaxEnt and empirical entropy}

The direct conditional MaxEnt framework models the activity of a binary output neuron by the least-structured conditional distribution that matches the output rate and the pairwise output-input coactivities, yielding a logistic artificial neuron \cite{Jaynes1957,WainwrightJordan2008,Lynn2026},
\begin{equation}
  P_{\rm dir}(y=1\mid x)
  =
  \sigmoid\left(b+\sum_i w_i x_i\right),
  \label{eq:lynn_model}
\end{equation}
The model entropy is averaged over the empirical distribution of recorded input states,
\begin{equation}
  \Sdir
  =
  \E_{x\sim \Pemp}
  \left[
    h\left(P_{\rm dir}(y=1\mid x)\right)
  \right],
\end{equation}
where \(h(\cdot)\) is the Bernoulli entropy. Here, ``direct'' refers to additivity of the conditional log-odds in the selected inputs, with no explicit products such as \(x_ix_j\).

Let \(\Stot\) be the marginal output entropy, \(\Sdir\) the conditional entropy of the fitted direct model, and \(\Strue\) the true conditional entropy. In synthetic data, where \(\Strue\) is known, we use
\begin{equation}
  \phidir
  =
  \frac{\Stot-\Sdir}{\Stot-\Strue}
\end{equation}
to measure the fraction of recoverable information captured by the direct model. In real data, \(\Strue\) is unknown, so direct entropy explained is a predictive summary under \(\Pemp(x)\), not a mechanistic decomposition. The same distinction applies more generally: entropy explained by a direct model is a statement about prediction under \(\Pemp(x)\); prediction of raw higher-order coactivities is a statement about averages over \(\Pemp(x)\); and fitted direct parameters are Lagrange multipliers for activity statistics, not causal synaptic parameters.

The predictive interpretation of this construction is straightforward. If \(\Sdir\) is small, the fitted direct model is close to the true conditional response on average over the sampled input distribution. The harder question relates to mechanism identification. The entropy calculation does not ask whether the same response would remain first-order under a different state distribution, nor whether omitted sufficient statistics are correlated with the included direct statistics. The analyses below examine that identifiability question, rather than predictive accuracy.

\section{Information geometry of restricted conditional MaxEnt fits}

\subsection{Omitted sufficient statistics shift the fitted direct parameters}

The identifiability argument below uses standard information-geometric notions of \(I\)-projection onto an exponential family \cite{Csiszar1975,Amari2016}. In other words, the restricted MaxEnt model does not ask whether omitted statistics exist. It asks whether the effect of those omitted statistics has a component that cannot be represented by the included statistics under the sampled input distribution.

Let \(F(y,x)\) denote the included sufficient statistics and \(G(y,x)\) the omitted statistics. In the direct conditional MaxEnt model,
\begin{equation}
  F(y,x)=(y,yx_1,\ldots,yx_n),
\end{equation}
such that Eq.~\ref{eq:lynn_model} is the corresponding conditional exponential family.

Suppose that the true conditional response also contains omitted statistics:
\begin{equation}
  P_{\theta_0,\lambda}(y\mid x)
  =
  \frac{
    \exp\{\theta_0^\top F(y,x)+\lambda^\top G(y,x)\}
  }{
    Z_{\theta_0,\lambda}(x)
  }.
\end{equation}
The restricted MaxEnt fit ignores \(G\) and chooses \(\hat\theta\) to match the \(F\)-moments of the true model. Define the conditional Fisher covariance blocks at \(\lambda=0\):
\begin{align}
  I_{FF}
  &=
  \E_{\Pemp(x)}
  \left[
    \Cov_{P_{\theta_0}(y\mid x)}(F,F\mid x)
  \right],\\
  I_{FG}
  &=
  \E_{\Pemp(x)}
  \left[
    \Cov_{P_{\theta_0}(y\mid x)}(F,G\mid x)
  \right].
\end{align}
Differentiating the moment-matching equations gives the local response of the restricted fit to a small omitted coefficient,
\begin{equation}
  \hat\theta-\theta_0
  =
  I_{FF}^{-1}I_{FG}\lambda
  +
  O(\|\lambda\|^2)
  .
  \label{eq:fisher_leakage}
\end{equation}
Thus, omitted statistics leak into fitted first-order MaxEnt parameters whenever \(I_{FG}\neq 0\). No leakage requires Fisher-orthogonality between the included and omitted sufficient statistics under the sampled input distribution. Equation~\ref{eq:fisher_leakage} is a local statement around \(\lambda=0\); the exact-absorption construction below gives the corresponding nonlocal identifiability limitation. The information-projection framework is shown in Supplementary Fig.~\ref{fig:supp_concept}a and a resulting empirical non-identifiability example is shown in Fig.~\ref{fig:concept}.

The residual detectable by the restricted family is governed by the Schur complement
\begin{equation}
  I_{GG\mid F}
  =
  I_{GG}-I_{GF}I_{FF}^{-1}I_{FG}.
\end{equation}
Locally,
\begin{equation}
  \KL(P_{\theta_0,\lambda}\Vert P_{\hat\theta})
  =
  \frac12\lambda^\top I_{GG\mid F}\lambda
  +
  O(\|\lambda\|^3).
\end{equation}
Therefore, a small residual KL or small \(\Sdir\) can bound only the component of omitted structure that remains after projecting out \(F\). It does not bound the absorbed component unless one adds a separate lower bound on the Schur-complement curvature \(I_{GG\mid F}\).

\subsection{Exact absorption on restricted support}

The strongest non-identifiability case occurs when, on the observed input support,
\begin{equation}
  G(y,x)=A^\top F(y,x)+c(x).
\end{equation}
Because \(c(x)\) cancels in the conditional normalization \(Z(x)\),
\begin{equation}
  \theta^\top F+\lambda^\top G
  =
  (\theta+A\lambda)^\top F+\lambda^\top c(x)
\end{equation}
defines the same \(P(y\mid x)\) as a restricted model with shifted first-order parameters. Thus, a restricted MaxEnt fit can be exact even when the true representation contains an omitted statistic with \(\lambda\neq 0\). Observational data from the same support cannot distinguish these descriptions. The geometric picture is illustrated in Supplementary Fig.~\ref{fig:supp_concept}b.

This exact case is a limiting example, but the realistic case is partial absorption: \(G-A^\top F\) can have small weighted variance under \(\Pemp(x)\) and much larger variance under a different input distribution. In this case, the restricted model can be an excellent in-distribution predictor while failing a reweighted or cross-context test. The state-reweighting analyses below are designed to expose this partial-absorption case.

For example, if a recorded neuron satisfies \(x_3=x_1x_2\) on the observed support, then
\begin{equation}
  \logit P(y=1\mid x)=Kx_1x_2
\end{equation}
is indistinguishable from
\begin{equation}
  \logit P(y=1\mid x)=Kx_3.
\end{equation}
The first-order model would be an excellent predictor, but it would not identify whether the mechanism is a direct dependence on \(x_3\), a conjunction of \(x_1\) and \(x_2\), or an unobserved common cause.
This is not meant as a literal biological model. It is a proof of principle: perfect first-order prediction on restricted support does not identify the absence of an omitted higher-order mechanism.

\subsection{A closed-form coskewness identity for sparse binary inputs}

Express the activity of an input as \(x_i=\mu_i+\delta_i\), with \(\mu_i=\E[x_i]\) so that \(\delta_i=x_i-\mu_i\) is mean-zero. A pairwise product decomposes as
\begin{equation}
  x_i x_j
  =
  \mu_i\mu_j+\mu_i\delta_j+\mu_j\delta_i+\delta_i\delta_j.
  \label{eq:pair_decomp}
\end{equation}
The first three terms are constant or first-order. The centered product \(\delta_i\delta_j\) is identifiable as a second-order feature only to the extent that it is orthogonal to the first-order span under the input distribution. Its projection onto \(\delta_i\) is
\begin{equation}
  \Cov(\delta_i\delta_j,\delta_i)
  =
  \E[\delta_i^2\delta_j].
  \label{eq:coskew_general}
\end{equation}
This centered-product projection is the quantity relevant to nontrivial interaction leakage. For centered symmetric continuous inputs, the centered product is orthogonal to first-order features, so this coskewness leakage channel vanishes. Raw non-centered products can still carry first-order information through their means, because Eq.~\ref{eq:pair_decomp} contains constant and linear terms. Therefore, when the simulations are intended to test coskewness leakage rather than trivial mean leakage, the ground-truth interaction is defined using centered products.
For binary inputs \(x_i\in\{0,1\}\), with \(p_i=\E[x_i]=\mu_i\),
\begin{equation}
  \delta_i^2=(1-2p_i)\delta_i+p_i(1-p_i),
\end{equation}
so
\begin{equation}
  \E[\delta_i^2\delta_j]
  =
  (1-2p_i)\Cov(x_i,x_j)
  .
  \label{eq:binary_coskew}
\end{equation}
In sparse binarized data, \(p_i\ll 1/2\), so \(1-2p_i\approx 1\). Ordinary input covariance is then converted almost directly into a coskewness channel by which higher-order input features project onto first-order features (Supplementary Fig.~\ref{fig:supp_concept}c). This gives a concrete masking mechanism: the amount of apparent direct attribution can be determined by the input representation and input covariance, even when the underlying response generator is unchanged.

Binarization has a second consequence. Because \(x^n=x\) for a binary variable with \(n\geq1\), any single-input nonlinearity collapses into an affine function on \(\{0,1\}\):
\begin{equation}
  f(x_i)=f(0)+\big[f(1)-f(0)\big]x_i.
\end{equation}
Thus, threshold-like or saturating single-input effects cannot be distinguished from a direct weight after binarization. Multi-input interactions remain possible on the binary cube, but Eq.~\ref{eq:binary_coskew} shows that sparse correlated binary activity can also mask part of those interactions by projecting them into first-order features.

This is also the centered version of Eq.~\ref{eq:fisher_leakage}. For \(F_i=yx_i\) and \(G_{jk}=yx_jx_k\), the uncentered Fisher cross-covariance is
\begin{equation}
  I_{i,jk}
  =
  \E_{\Pemp(x)}
  \left[
    p(x)(1-p(x))x_i x_jx_k
  \right].
\end{equation}
Let \(v(x)=p(x)(1-p(x))\). Projecting \(G_{jk}\) onto the \(v(x)\Pemp(x)\)-weighted orthogonal complement of the constant and first-order span removes the constant and linear terms in Eq.~\ref{eq:pair_decomp}. The remaining cross-term is
\begin{equation}
  \E_{\Pemp(x)}
  \left[
    p(x)(1-p(x))\delta_i\delta_j\delta_k
  \right],
\end{equation}
a response-variance-weighted coskewness. If \(v(x)\) is approximately constant over the relevant local states, this becomes the input-only three-input coskewness \(\E[\delta_i\delta_j\delta_k]\), the multi-index generalization of the matched-index projection in Eq.~\ref{eq:coskew_general} (which it recovers when \(k=i\)). In sparse CA1 tables \(v(x)\) is not literally constant, so we do not use Eq.~\ref{eq:binary_coskew} as a numerical estimator of Fisher leakage. Instead, the empirical feature-leakage score in Methods computes the actual weighted projection under each local \(q(x)\). The practical consequence is the same: the conditional response can receive very different first-order attribution under different input-state distributions.

\section{Ground-truth simulations}

\subsection{Apparent direct attribution depends on representation and correlation}

The representation simulations use a purely higher-order generator with zero linear drive. With independent inputs, the direct model explains none of the recoverable information by construction, so any positive direct attribution is a masking artifact created by input distribution, covariance, or representation. Inputs are generated from a one-factor latent model with tunable correlation \(\rho\), then represented as binary, Gaussian, uniform, lognormal, or other distributions. Two models are examined on held-out data: the direct model \([1,x]\) and the interaction model \([1,x,x_ix_j]\). The key metric is
\begin{equation}
  \phidir
  =
  \frac{\Stot-\Sdir}{\Stot-\Strue}
  =
  \frac{\Idir}{\Itrue},
\end{equation}
the fraction of recoverable information captured by the direct model, where \(\Idir=\Stot-\Sdir\) and \(\Itrue=\Stot-\Strue\) are the recoverable information captured by the direct and the correctly specified models.

\begin{figure}[tbp]
  \centering
  \includegraphics[width=\textwidth]{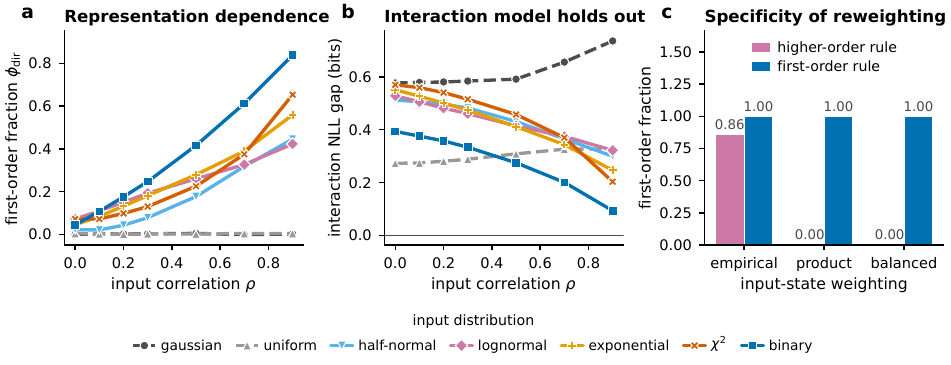}
  \caption{
  Ground-truth controls for distribution-sensitive first-order attribution. (a) A pure higher-order generator with zero linear drive gives very different first-order entropy attribution depending only on input representation and correlation. Seven representations are shown: binary correlated inputs can make the direct model capture much of the recoverable information (\(\phidir\approx 0.84\) at \(\rho=0.9\)); skewed positive representations (chi-squared, exponential, half-normal, lognormal) give intermediate attribution between the symmetric and binary extremes; and symmetric Gaussian and uniform representations with centered interaction features give \(\phidir\approx 0\) (Supplementary Table~\ref{tab:phi_direct_sweep}). (b) The interaction model remains predictively useful on held-out data across the same conditions and representations (colors as in a); positive values mean lower held-out negative log likelihood for the interaction model than for the direct model. (c) The reweighting control is specific: it removes apparent first-order attribution for a higher-order response but preserves a genuine first-order response.
  }
  \label{fig:main_controls}
\end{figure}

The same nonlinear generator yields very different apparent first-order attribution depending only on representation and correlation (Fig.~\ref{fig:main_controls}a). With centered interaction features, symmetric Gaussian and uniform representations give, as expected, \(\phidir\approx 0\) across the correlation sweep. For binary inputs, \(\phidir\) rises with correlation and reaches approximately \(0.84\) at \(\rho=0.9\), while skewed positive representations give intermediate values between the symmetric and binary extremes (half-normal \(\approx 0.45\), lognormal \(\approx 0.42\), exponential \(\approx 0.56\), and chi-squared \(\approx 0.65\) at \(\rho=0.9\); Supplementary Table~\ref{tab:phi_direct_sweep}). The interaction model remains predictive in all conditions, and the held-out negative log-likelihood (NLL) gap between direct and interaction models remains positive (Fig.~\ref{fig:main_controls}b). Thus, the nonlinear signal is recoverable; the direct model captures it only when the input geometry projects interactions into first-order features.

A complementary greedy-dictionary analysis and a larger \(K=320\) check gave the same qualitative result: the direct attribution depends strongly on representation, while interaction features continue to improve held-out likelihood. We keep those robustness checks in the analysis package rather than the main figure, because the central point is already made by the representation sweep and held-out interaction gap.

\FloatBarrier

The following controls are more stringent because they hold the conditional response table \(P(y\mid x)\) fixed and change only the input-state weights \(q(x)\). They ask whether first-order attribution is a stable property of the response function or whether it changes when the same response is evaluated under a different distribution of sampled input states.

\subsection{Reweighting separates response mechanism from state weighting}

For small input sets, the response table \(r(x)=P(y=1\mid x)\) can remain fixed while the input-state weights \(q(x)\) are changed. We evaluate the same response table under
\begin{equation}
  \qemp(x),\qquad
  \qprod(x)=\prod_i P(x_i),\qquad
  \qbal(x)=2^{-k}.
\end{equation}
For an order-\(m\) logistic model, define the controlled available-entropy attribution
\begin{equation}
  F_m(q)
  =
  \frac{S_0(q)-S_m(q)}
       {S_0(q)-\Ssat(q)}.
  \label{eq:Fm}
\end{equation}
If a response is genuinely first-order, \(F_1(q)\) should remain high as \(q\) changes. If first-order success depends on empirical input geometry, \(F_1(\qemp)\) can be high while \(F_1(\qbal)\) drops.

\begin{figure}[tbp]
  \centering
  \includegraphics[width=\textwidth]{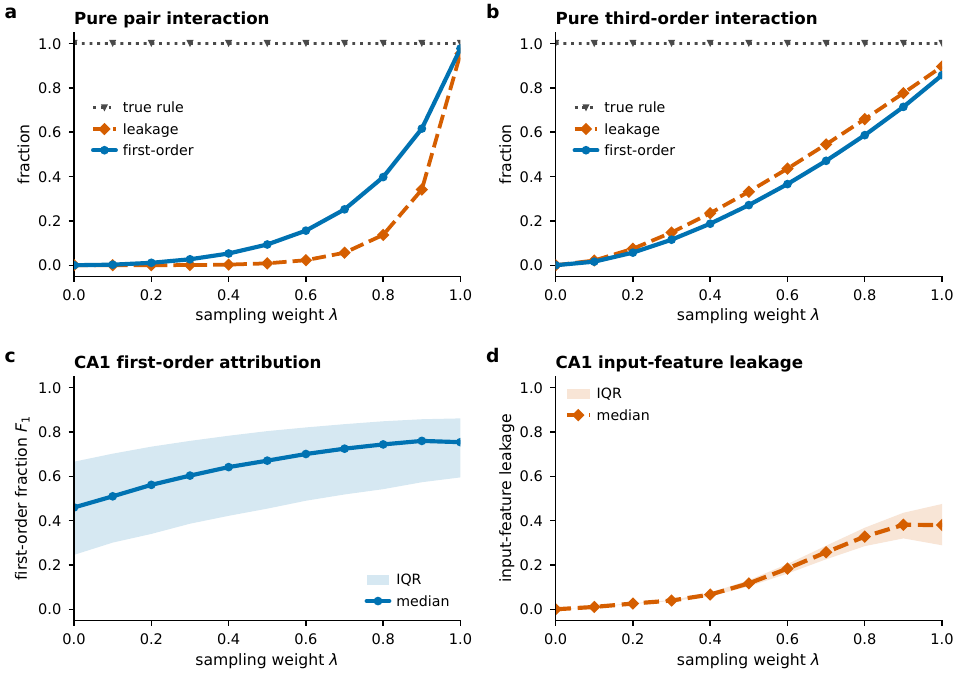}
  \caption{
  State-weight tuning with the response table \(P(y\mid x)\) held fixed; the \(x\)-axis \(\lambda\) sweeps the input-state distribution from balanced (\(\lambda=0\)) toward a leakage-prone/empirical distribution (\(\lambda=1\)). We use \emph{cube} (equivalently \emph{local cube}) as a shorthand for the conditional response table of a single binary output \(y\) to a chosen triple of its binary inputs \((x_1,x_2,x_3)\): the three inputs take \(2^3=8\) joint states---the eight corners of a cube in \(\{0,1\}^3\)---and the table records \(P(y{=}1\mid x_1,x_2,x_3)\) at each corner. A \emph{CA1 cube} is such a table estimated from the mouse hippocampal CA1 recording, with the output \(y\) and the three inputs \(x_1,x_2,x_3\) all recorded CA1 neurons. \textbf{a,b} For pure second- and third-order synthetic response tables, the first-order attribution (blue) rises as \(\lambda\) increases, while the correctly specified higher-order model (gray dotted ``true rule'') stays fully explanatory; the orange dashed curve is the accompanying projection of higher-order input features onto first-order features. \textbf{c,d} For estimated CA1 cubes---each cube's response table held fixed while only the weighting of its \(8\) input states is swept---moving the input-state distribution from balanced toward the empirical CA1 distribution increases both the first-order entropy attribution \(F_1\) (c) and the higher-order-to-first input-feature leakage (d); lines are medians and bands are the interquartile range across cubes. The apparent amount of first-order explanation therefore changes because \(P(x)\) changes, not because \(P(y\mid x)\) changes.
  }
  \label{fig:state_tuning}
\end{figure}

In ground-truth simulations, the control behaves selectively (Fig.~\ref{fig:main_controls}c). For a pure third-order response, first-order attribution is high under a leakage-prone empirical weighting but disappears after product or balanced reweighting, while the third-order model remains correct. For a genuine additive first-order response on the same restricted input manifold, the first-order model remains accurate under all three weightings. Reweighting therefore does not simply erase first-order structure; when the response is genuinely additive, the first-order fit remains stable. The diagnostic also separates fundamentally different response types tested on the same input grid: a pure XOR-like interaction fails under balanced reweighting, an asymmetric AND-like interaction fails under empirical reweighting, and a genuine additive response is preserved under both (Fig.~\ref{fig:concept}). The continuous version of this control, in which the state weights are interpolated from balanced to a leakage-prone endpoint, makes the dependence on \(P(x)\) explicit (Fig.~\ref{fig:state_tuning}). Adding inputs does not remove the effect: under a leakage-prone empirical distribution, empirical first-order attribution for pure higher-order generators \emph{grows} with the number of inputs \(K\) (from \(0.36\) to \(0.63\) for a second-order generator and from \(0.55\) to \(0.87\) for a third-order generator over \(K\in[3,10]\)), because the number of higher-order features that can serve as first-order proxies grows combinatorially, while balanced attribution stays near zero (Supplementary Fig.~\ref{fig:supp_scaling}).

\FloatBarrier

\section{Distribution-sensitive attribution in CA1 and visual cortex}

The real-data analyses should be considered as local identifiability diagnostics. They do not attempt to reproduce the full high-dimensional greedy model neuron-by-neuron. Instead, they exploit small well-sampled input subsets for which all binary states can be enumerated, allowing us to hold the estimated response table \(P(y\mid x)\) fixed while changing \(P(x)\). Concretely, a \emph{local cube} is one such three-input response table: the three binary inputs span \(2^3=8\) states (the corners of a cube in \(\{0,1\}^3\)), each of which is observed often enough to estimate \(P(y\mid x)\) and to reweight exactly. When the output \(y\) and its three inputs are all neurons from the mouse hippocampal CA1 recording, we refer to this table as a \emph{CA1 cube}. This gives the minimal control needed to separate response form from state weighting.
These local cubes are not a random survey of all possible CA1 triplets; they are diagnostic subsets selected to test identifiability in regimes where first-order prediction and higher-order-to-first feature leakage are both measurable.

\subsection{CA1 occupies a leakage-prone sparse binary regime}

The publicly available binarized hippocampal CA1 activity matrix released with \cite{Lynn2026} contains \(1{,}485\) neurons and \(70{,}338\) bins. The global activity probability is \(0.0185\), the median neuron activity probability is \(0.0127\), and the median binary factor \(1-2p_i\) is \(0.9746\). The sparse-binary coskewness factor in Eq.~\ref{eq:binary_coskew} is therefore almost maximally active for a typical neuron, placing the dataset in the leakage-prone regime predicted by the framework.

\begin{figure}[tbp]
  \centering
  \includegraphics[width=\textwidth]{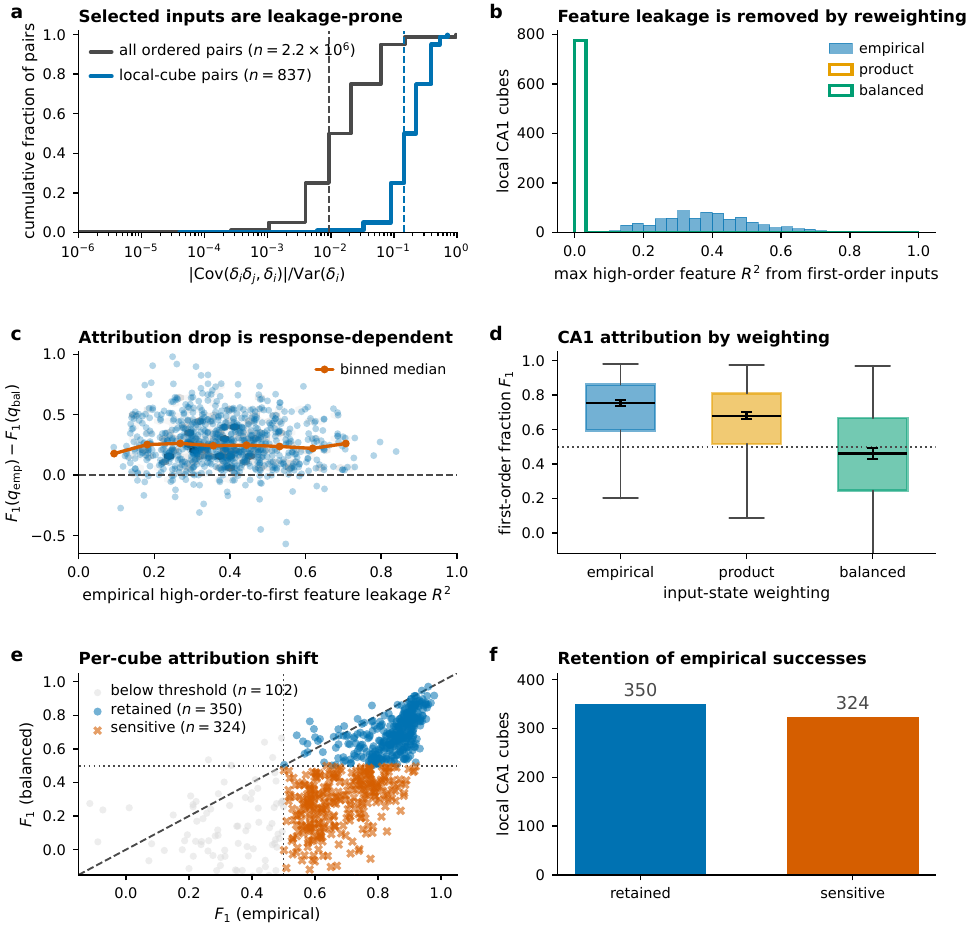}
  \caption{
  Local CA1 reweighting diagnostics. \textbf{a} Ordered coskewness-projection coefficients are larger for input pairs selected into the local-cube analysis than across all CA1 pairs, placing the diagnostic subsets in a leakage-prone regime. \textbf{b} Centered higher-order input features have substantial first-order projection under empirical weights; product and balanced weights remove most of that feature leakage (\(n=776\) cubes). \textbf{c} The empirical-to-balanced \(F_1\) drop is not explained by input-feature leakage alone (Spearman \(\rho=-0.05\)), showing that the response table matters; shading is point density. \textbf{d} Estimated local CA1 response tables show high first-order attribution under empirical weights that drops under product and balanced weights (boxplots over \(n=776\) cubes; medians with bootstrap 95\% CIs). \textbf{e} Per-cube first-order attribution under empirical versus balanced weights; each point is one local response table, colored as retained (\(n=350\)) or distribution-sensitive (\(n=324\)) of the \(674\) empirical successes, or gray if below the empirical-success threshold (\(n=102\), of \(776\) cubes total). \textbf{f} Counts of retained versus distribution-sensitive tables among the \(674\) successes.
  }
  \label{fig:ca1_entropy}
\end{figure}

Across all \(\binom{1485}{2}=1{,}101{,}870\) neuron pairs, only \(32\%\) have positive Pearson correlation; the mean correlation is \(0.002\), the 95th percentile is \(0.047\), and the 99th percentile is \(0.105\). The bulk distribution is therefore weakly coupled. However, selected local input sets are enriched for leakage-prone pairs: in well-sampled local cubes, selected input pairs have the median Pearson correlation \(0.173\) and the median projection coefficient \(0.146\), compared with much smaller values across all ordered pairs (Fig.~\ref{fig:ca1_entropy}a; the median absolute coskewness projection across all pairs is \(0.009\)). Here, the projection coefficient is the normalized regression slope \(\mathrm{Cov}(\delta_i\delta_j,\delta_i)/\mathrm{Var}(\delta_i)\), i.e.\ the numerator of Eq.~\ref{eq:binary_coskew} divided by \(\mathrm{Var}(\delta_i)\), so it is directional and reported per ordered pair. Centered higher-order input features have substantial first-order projection under empirical weights, and product or balanced state weights remove most of that projection (Fig.~\ref{fig:ca1_entropy}b). The size of the entropy drop is response-dependent, not a direct consequence of input geometry alone (Fig.~\ref{fig:ca1_entropy}c). Additional activity-rate diagnostics are shown in Supplementary Fig.~\ref{fig:supp_ca1_coskew}; together these analyses motivate the entropy reweighting test but do not by themselves prove the existence of a specific higher-order mechanism.

\subsection{Local CA1 entropy attribution changes after state reweighting}

For \(776\) well-sampled three-input CA1 cubes across \(40\) targets, we estimated \(P(y\mid x)\) locally and recomputed entropy attribution under empirical, product, and balanced input-state weights. This local diagnostic complements high-dimensional greedy modeling by focusing on subsets in which all \(2^3\) input states can be enumerated and exactly reweighted. These cubes are deliberately selected to be leakage-prone -- inputs are chosen for high output coactivity, so the subset is enriched for exactly the sampling geometry under which absorption can occur. They are therefore a diagnostic stress test of how large the effect can be when the direct-MaxEnt framework is applied, \emph{not} a representative estimate of the fraction of CA1 responses that are mechanistically higher-order.

The local result is mixed in an important way. Under empirical weights, first-order attribution is high (median \(F_1(\qemp)=0.754\)). We treat the \emph{product}-marginal control as the primary, conservative de-leaking statistic and the \emph{balanced} reweighting as a stress-test endpoint. Under the product control the median drops modestly to \(0.679\); under the balanced stress test it drops to \(0.459\). Among local tables with \(F_1(\qemp)\ge 0.5\), \(350/674\) remain above threshold after balanced reweighting, while \(324/674\) do not (retained fraction \(0.519\), bootstrap \(95\%\) CI \(0.482\)--\(0.556\)). Thus, within these leakage-enriched diagnostic tables, roughly half of those that look first-order under empirical weights are distribution-sensitive.

This split is not a finite-sample or smoothing artifact. We ran a matched additive-surrogate null: for each cube we fit the best additive (first-order) response, simulated spike counts at the observed per-state counts and cross-validation folds, and pushed each surrogate through the identical smoothing, reweighting and classification pipeline (\(20\) simulations per cube). Genuinely additive tables are essentially never misclassified as distribution-sensitive: \(0\) of \(13{,}480\) additive-surrogate tables crossed the threshold (\(95\%\) upper bound \(0.02\%\) by the rule of three), so the expected distribution-sensitive count under the null is \(\approx 0\) of \(674\). The observed \(324/674\) (\(48\%\)) is therefore far above what estimation noise produces (Supplementary Fig.~\ref{fig:supp_surrogate}).

To check whether the effect depends on selecting inputs by their output coactivity, we also computed an unselected base-rate control: input triplets were chosen only by their mutual co-occurrence (so the \(2^3\) table can be estimated) and paired with randomly chosen targets, with no output-based selection. Of \(3{,}000\) evaluated tables, \(350\) met the empirical-success criterion. Among these empirical-success tables the distribution-sensitive fraction was \(0.74\) (\(259/350\); \(95\%\) CI \(0.69\)--\(0.79\)), with median \(F_1\) dropping from \(0.638\) under empirical weights to \(0.316\) under balanced weights, and \(95\%\) (CI \(92\)--\(97\%\)) were conditionally non-additive by the \(\Delta_{ij}\) surrogate test. Removing the output-coactivity selection therefore does not remove the effect. Because the unselected analysis is still conditioned on well-sampled input triplets and on empirical-success tables, we use it as a robustness check rather than a population prevalence estimate.

This result separates two components of empirical first-order entropy attribution: robust first-order predictive structure and attribution that depends on the empirical input-state distribution.

\FloatBarrier

\subsection{Visual-cortex recordings provide a secondary generality check}

Publicly available mouse visual-cortex natural-image and spontaneous recordings provide a second test of the framework. These recordings have many cells but far fewer time bins than CA1, so exact three-input cubes are often under-sampled. We therefore repeated the cross-dataset analysis with two-input state tables. The reweighting control still bears on higher-order coactivities, because an omitted pair interaction \(x_ix_j\) is the conditional response feature associated with the second-order coactivity \(\langle yx_ix_j\rangle\) (throughout, we index a coactivity by its number of co-active inputs \(m\), so \(\langle yx_ix_j\rangle\) has order \(m=2\) and \(\langle yx_ix_jx_k\rangle\) has order \(m=3\)).

The visual-cortex effects are smaller than CA1 and more consistent with robust first-order structure. For natural-image responses, median local \(F_1\) drops from \(0.602\) under empirical weights to \(0.511\) under balanced weights, with \(107/182\) local successes retained. For spontaneous activity, the median drops from \(0.653\) to \(0.579\), with \(120/173\) retained. A matched-population control, downsampling V1 to \(N=1{,}485\) neurons and recomputing CA1 with the same two-input local tables, shows that the cross-dataset comparison is not simply an artifact of V1 having many more recorded neurons. This control matches neuron count but not recording duration, so we treat the V1 results as a secondary check rather than the main empirical claim.
The cross-dataset summary is shown in Supplementary Fig.~\ref{fig:supp_cross_dataset} and the population-matched control in Supplementary Fig.~\ref{fig:supp_matched_population}. They are secondary to the CA1 analysis because the visual recordings support only smaller local tables, but they show that the distribution-sensitive component is not unique to CA1.

\section{Raw coactivities are not conditional interaction tests}

A complementary class of tests of direct-dependency models asks whether they predict raw second-, third-, fourth-order, and time-delayed coactivity rates. In binarized neural recordings these raw coactivity scatters often pass with high accuracy: reported direct models predict \(99.85\%\) of second-order coactivities \(\langle yx_ix_j\rangle\) and \(99.89\%\) of third-order coactivities \(\langle yx_ix_jx_k\rangle\) within experimental errors in CA1 \cite{Lynn2026}. These are strong descriptive results. The issue is that raw coactivities are not conditional interaction coefficients. For example, the predicted second-order coactivity is
\begin{equation}
  \langle yx_ix_j\rangle_{\rm pred}
  =
  \E_{\Pemp(x)}
  \left[
    P_{\rm dir}(y=1\mid x)x_ix_j
  \right].
\end{equation}
This expectation depends explicitly on \(\Pemp(x)\). If the observed state distribution makes \(x_ix_j\) predictable from first-order features, then a first-order model can predict \(\langle yx_ix_j\rangle\) even when the conditional response table contains an interaction term. An analogous issue appears for \(\langle yx_ix_jx_k\rangle\), \(\langle yx_ix_jx_kx_\ell\rangle\), and time-delayed coactivities: these are raw averages over the sampled state distribution, not direct estimates of conditional interaction coefficients.

Raw coactivity prediction combines two ingredients: the fitted conditional response \(P(y\mid x)\) and the input-state distribution \(P(x)\). A useful control is therefore to hold the response table fixed and recompute the same quantities under alternative state weights. If the first-order model remains successful, the result is consistent with robust first-order response structure. If the success weakens, the coactivity prediction depended partly on the sampling distribution.

A simple simulation illustrates the issue. We simulated eight binary inputs grouped into four pairs. The ground-truth response contains only pairwise XOR-like terms: each pair contributes to the log odds when its state is \(10\) or \(01\), but not when it is \(00\) or \(11\). Under a balanced input distribution, these terms are not additive. Under a rare-corner input distribution in which the \(11\) state has probability \(0.005\) for each pair, the same conditional response is close to an additive OR-like rule on the sampled states. A first-order MaxEnt model then captures \(0.91\) of the recoverable entropy under the rare-corner distribution but approximately \(0\) under balanced state weights.

\begin{figure}[tbp]
  \centering
  \includegraphics[width=\linewidth]{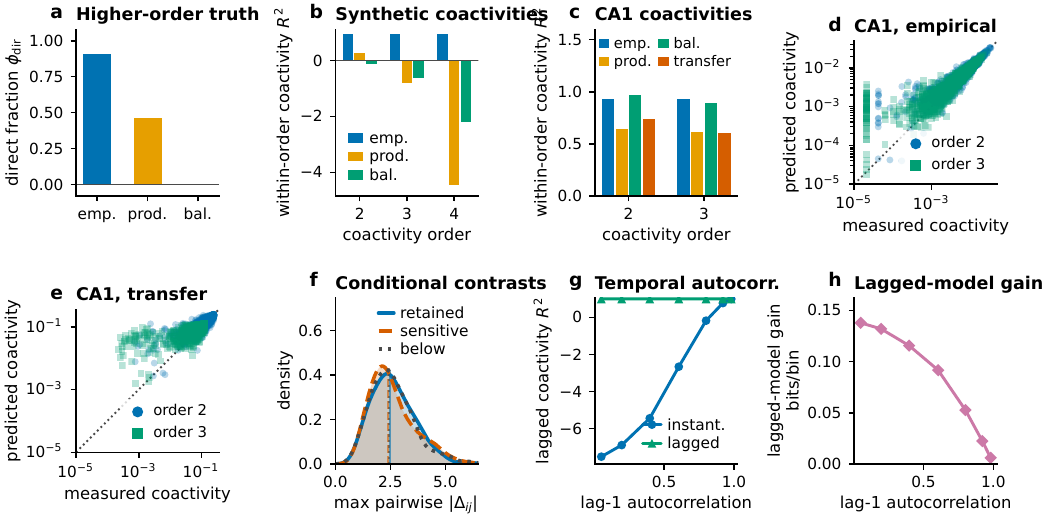}
  \caption{
  Raw coactivity prediction is a descriptive test, not an interaction-identification test. \textbf{a,b} In a known higher-order simulation, rare-corner sampling makes a first-order model explain entropy and predict raw higher-order coactivities; product and balanced state weights expose the mismatch. Negative \(R^2\) in (b) indicates performance worse than the within-order mean-coactivity baseline. \textbf{c--e} In local CA1 response tables, raw coactivity \(R^2\) changes across empirical, product-refit, balanced-refit and empirical-to-balanced transfer controls; the scatter plots show the empirical and transfer cases in raw coactivity space. \textbf{f} Conditional pairwise logit contrasts remain large in retained, distribution-sensitive, and below-threshold local tables, showing that raw coactivity prediction and conditional additivity are different questions. \textbf{g,h} In a known lagged generator, instantaneous features predict delayed coactivities when present and past inputs are highly autocorrelated; the mechanism test is the held-out comparison in (h), where the correctly lagged model improves negative log likelihood across the autocorrelation sweep.
  }
  \label{fig:nature_fig4_control}
\end{figure}

The entropy and raw coactivity metrics show the same issue (Fig.~\ref{fig:nature_fig4_control}a,b). Under the rare-corner distribution, the direct model captures \(0.91\) of the recoverable information, and first-order predictions of second-, third-, and fourth-order coactivities have \(R^2=0.99,0.98,0.96\), respectively. Here \(R^2\) is the coefficient of determination between the first-order model's \emph{predicted} coactivities \(\langle yx_{i_1}\cdots x_{i_m}\rangle_{\rm pred}=\E_{P(x)}[P_{\rm dir}(y=1\mid x)\,x_{i_1}\cdots x_{i_m}]\) and the corresponding \emph{measured} coactivities \(\E_{P(x)}[y\,x_{i_1}\cdots x_{i_m}]\), taken across all within-order terms (the \(\binom{K}{m}\) input \(m\)-tuples for order \(m\)). Under product weights, the direct entropy fraction drops to \(0.46\), and the corresponding coactivity \(R^2\) values fall to \(0.30,-0.81,-4.48\). Under balanced weights, the direct fraction is approximately zero and the corresponding within-order coactivity \(R^2\) values are \(-0.15,-0.66,-2.22\). Negative \(R^2\) values mean that the first-order predictions are worse than using the within-order mean coactivity as a baseline. In this example, a raw coactivity metric looks excellent even though the ground truth is explicitly higher-order.

The CA1 local tables give the real-data analogue, with an important distinction between refit and transfer controls. Using the empirical local state weights, the first-order model predicts local second- and third-order coactivities with \(R^2=0.93\) and \(0.94\), respectively. If the best first-order model is refit under product local weights, the corresponding \(R^2\) values drop to \(0.64\) and \(0.62\). If the best first-order model is refit under balanced local weights, raw coactivity prediction remains high for local second-order coactivities and lower, but still substantial, for local third-order coactivities (\(R^2=0.97\) and \(0.90\)). This reinforces the point that raw coactivity prediction is not a direct conditional-interaction test. In contrast, if the empirical first-order fit is transferred to balanced local weights, the corresponding \(R^2\) values drop to \(0.75\) and \(0.61\), showing that the empirical fit itself depends on state weighting. The primary identifiability statistic is the refit entropy control: the local first-order entropy fraction drops from a median of \(0.754\) under empirical weights to \(0.679\) under product weights and \(0.459\) under balanced weights. The empirical state distribution is also strongly uneven: across the \(776\) local CA1 cubes, the median least-sampled state has probability \(0.0031\), only \(2.5\%\) of its balanced weight, and \(96.4\%\) of cubes have a least-sampled state below \(5\%\) of balanced weight. Thus, the rare-corner simulation is not meant as a random model of CA1; it isolates a sampling pattern that is common in the selected local CA1 diagnostic tables.

The conditional log-odds contrast confirms that raw coactivity success is not the same as conditional additivity. The median maximum pairwise conditional contrast is \(2.44\) across local cubes. It remains large in both retained and distribution-sensitive successes: the corresponding medians are \(2.50\) and \(2.38\). The reweighting split is not simply separating additive from non-additive tables; rather, it separates response tables whose first-order attribution is stable across state distributions from those whose attribution depends on empirical sampling (Fig.~\ref{fig:nature_fig4_control}c--f). The log-odds contrast is not used to classify local tables as biologically nonlinear; it is used to show that raw coactivity prediction and conditional additivity are different statistical questions. This is the key empirical conclusion: the CA1 data contain strong first-order predictive structure, but the amount attributed to first-order direct dependencies by empirical entropy and raw coactivity tests is distribution-sensitive.

The more direct local test is a conditional odds-ratio contrast. For two binary inputs,
\begin{align}
  \Delta_{ij}
  ={}&
  \logit P(y=1\mid x_i=1,x_j=1)
  -\logit P(y=1\mid x_i=1,x_j=0)
  \notag\\
  &-\logit P(y=1\mid x_i=0,x_j=1)
  +\logit P(y=1\mid x_i=0,x_j=0).
  \label{eq:odds_contrast}
\end{align}
For an additive logistic model, \(\Delta_{ij}=0\). Unlike raw coactivity prediction, this statistic targets the conditional response table and is independent of the input-state distribution \(P(x)\); it therefore answers a different question from the reweighting control (which asks whether entropy \emph{attribution} depends on \(P(x)\)), and the two are complementary. We calibrate it with the same additive-surrogate machinery: a finite-sample, smoothed table from a genuinely additive response still has a nonzero observed \(\max_{ij}|\Delta_{ij}|\), so for each cube we build the null distribution of \(\max_{ij}|\Delta_{ij}|\) from additive surrogates at the observed per-fold counts and call a table conditionally non-additive when its observed value exceeds the per-cube null \(95\)th percentile. In the selected leakage-enriched cubes the observed median \(\max_{ij}|\Delta_{ij}|\) is \(2.44\) versus an additive-null median of \(0.44\), and nearly all of these tables exceed their additive null (\(673/674\) empirical-success cubes; \(99.9\%\), \(95\%\) CI \(99.6\)--\(100\%\)). Within these diagnostic subsets the local response tables therefore carry genuine, \(P(x)\)-independent interaction structure, complementary to the distribution-sensitivity of their first-order entropy attribution; as above, this is a statement about the selected, high-coactivity tables, not a prevalence estimate over all CA1 triplets.

An analogous problem appears for time-delayed coactivities. If \(x_i(t)\) is correlated with \(x_i(t-\tau)\), instantaneous sufficient statistics can predict time-delayed coactivities even when the ground truth contains lagged terms (Supplementary Fig.~\ref{fig:supp_visual_temporal}). We simulated a known lagged generator in which \(y_t\) depends on \(x_i(t-1)\), not on \(x_i(t)\), across a sweep of input autocorrelations. As lag-one autocorrelation increases, the instantaneous model increasingly passes the lagged-coactivity metric, reaching \(R^2=0.996\) at autocorrelation \(0.98\), even though the ground truth is purely lagged. Across the same sweep, the correctly lagged model recovers the delayed coactivities and improves held-out negative log likelihood, with gains ranging from \(0.006\) to \(0.138\) bits per sample (Fig.~\ref{fig:nature_fig4_control}g,h). Prediction of time-delayed coactivities is therefore also a descriptive test, not an identification test for instantaneous mechanisms.

\FloatBarrier

\section{Discussion}

The main point of the framework is to separate two questions that are often treated as one: (1) whether a direct (i.e., linear) model predicts well, and (2) whether that success identifies the underlying response mechanism. The most useful tools are state reweighting, conditional log-odds contrasts, and temporal model comparison. Reweighting asks whether the same response table still looks first-order after the state distribution changes. The log-odds contrast tests additivity directly, rather than through raw coactivity averages. Temporal model comparison addresses the analogous problem in time, where instantaneous variables can absorb lagged mechanisms. Together with the information-projection calculation (Eq.~\ref{eq:fisher_leakage}) and the binary coskewness identity (Eq.~\ref{eq:binary_coskew}), these tools give direct-dependency MaxEnt analyses a clearer interpretive scope.

The analyses make clear why high predictive accuracy is not the same thing as mechanistic inference. Entropy explained by direct dependencies, and prediction of raw higher-order or time-delayed coactivities, can arise when omitted statistics project onto the included first-order sufficient statistics. The relevant distinction is therefore not ``direct models predict'' versus ``direct models fail,'' but whether the first-order attribution remains stable after controlling the input-state geometry on which the MaxEnt average is computed. Applied to publicly available recordings, the framework finds that about half of local first-order successes in CA1 survive balanced reweighting, and about half are distribution-sensitive. In visual cortex, most local successes are more robust, with a smaller but still measurable distribution-sensitive component. This contrast should be read with care: the primary CA1 analysis uses three-input local tables while the visual-cortex analysis uses two-input tables, and three-input tables admit more higher-order features that can leak into first-order attribution, so part of the larger CA1 effect may reflect input-table order rather than region per se. The matched two-input, population-size-controlled comparison (Supplementary Fig.~\ref{fig:supp_matched_population}) is the cleaner cross-dataset test, and a measurable distribution-sensitive component remains in both. The qualitative point is that the control is not simply pushing every dataset away from a first-order interpretation: it separates the tables whose first-order attribution is stable from those whose attribution depends on the observed state distribution.

A related caution applies to weights and connectivity. A fitted MaxEnt weight is a Lagrange multiplier chosen to match activity statistics, not a causal synaptic parameter. Shared latent variables, recurrent dynamics, unobserved neurons, stimulus drive, and behavioral state can all create apparent direct dependencies \cite{Goris2014,Morrell2021,DasFiete2020,Macke2011}. None of this makes direct MaxEnt models uninformative. It means that the successful prediction of a direct model should be reported as prediction by first-order sufficient statistics under \(\Pemp(x)\), not as evidence that omitted mechanisms are absent. The same projection argument carries over to time and latent variables: if \(x(t)\) is correlated with \(x(t-\tau)\), instantaneous sufficient statistics can absorb lagged mechanisms; if a latent state drives both \(x_i\) and \(y\), fitted direct parameters can absorb common drive.

It also helps place standard controls in context. Random-input controls, train/test splits, and cross-validation address overfitting and finite-sample noise; they do not address whether omitted sufficient statistics are absorbed under the empirical input distribution. The state-reweighting control, the \(\Delta_{ij}\) contrast, and the temporal leakage comparison address that question explicitly. None of these tools replaces a full high-dimensional generative model when one is feasible, but each is computable on the small well-sampled local input subsets that direct MaxEnt analyses already produce, and each can be added to existing analyses with modest additional cost.

These results are consistent with two largely complementary regimes identified in the maximum-entropy population-modeling literature. The CA1 and visual-cortex recordings sit between the perturbative regime described by Roudi, Nirenberg, and Latham \cite{Roudi2009}, in which sparse, weakly coupled binary populations are well captured by pairwise models almost by construction, and the strongly correlated regime described by Schneidman et al.\ \cite{Schneidman2006,Tkacik2014,Ganmor2011}, in which weak pairwise correlations across many neurons combine into collective network states that pairwise MaxEnt nonetheless reproduces. Both regimes are compatible with high empirical first-order entropy attribution, and neither rules out higher-order interactions \emph{at the level of the conditional response}. The state-reweighting and \(\Delta_{ij}\) controls developed here are sensitive to the latter while remaining unaffected by the former, and, therefore, add an axis of identifiability information that the previous literature did not have to address explicitly.

Direct MaxEnt remains a useful description of binarized neural activity under the observed distribution of recorded states. The caution is interpretive: direct entropy explained and raw coactivity prediction are not mechanistic decompositions. The biological question of whether neurons integrate inputs additively or through higher-order, temporal, or latent mechanisms remains open and requires controls that separate \(P(y\mid x)\) from \(P(x)\), test conditional interactions directly, and compare instantaneous models with temporal and latent-variable alternatives.  

\section{Methods}

\subsection{Conditional MaxEnt and entropy notation}

All entropy quantities are measured in bits. For a finite local input table with states \(x\in\{0,1\}^k\), response table \(r(x)=P(y=1\mid x)\), and input-state weights \(q(x)\), the saturated conditional entropy is
\begin{equation}
  \Ssat(q)
  =
  \sum_x q(x)h(r(x)).
\end{equation}
The zero-input entropy is
\begin{equation}
  S_0(q)
  =
  h\left(\sum_x q(x)r(x)\right).
\end{equation}
For an order-\(m\) logistic table model with prediction \(p_m(x)\), the evaluated conditional entropy is
\begin{equation}
  S_m(q)
  =
  \sum_x q(x)h(p_m(x)).
\end{equation}
The controlled attribution fraction \(F_m(q)\) in Eq.~\ref{eq:Fm} is therefore the fraction of the available entropy reduction \(S_0(q)-\Ssat(q)\) captured by the order-\(m\) model under the specified input-state weights.
This fraction can become noisy when the available entropy denominator is very small. The local CA1 tables were therefore required to be well sampled in every state. When numerical fitting produced non-finite values or extreme fractions because of a near-zero denominator, values were clipped to \([-1.5,1.5]\) for storage and aggregation after recording the available entropy denominator. Across the \(776\) well-sampled local CA1 cubes, the empirical available entropy had median \(0.061\) bits and only \(8/776\) cubes below \(0.02\) bits; under balanced weights the median available entropy was \(0.254\) bits.

\subsection{Synthetic representation and robustness simulations}

The representation simulations used a pure higher-order generator with zero linear drive (parameters in Supplementary Table~\ref{tab:synthetic_parameters}). Inputs were generated by a one-factor latent model with correlation parameter \(\rho\), then transformed into the different representations summarized in Supplementary Table~\ref{tab:input_representations}: binary, Gaussian, uniform, lognormal and additional skewed distributions. The output was binary with conditional log-odds determined by centered pairwise products. Direct and interaction logistic models were fit on held-out data, and performance was measured in bits using \(\phidir=(\Stot-\Sdir)/(\Stot-\Strue)\), the corresponding interaction fraction, and the held-out NLL gap.

The greedy-dictionary simulations used the same ground-truth generator but allowed the feature-selection procedure to choose from direct inputs \(x_i\) and pairwise products \(x_ix_j\). This analysis asks whether nonlinear information that is missed or absorbed by the direct model is recovered once explicit interaction features are available. The \(K=320\) check repeats the entropy and held-out likelihood comparison in a larger input population.

\subsection{State reweighting}

For small input sets, we estimate or specify the conditional response table \(r(x)=P(y=1\mid x)\) and evaluate the same table under alternative state weights. Reweighting changes the measure over input states but not the estimated response table: for each local state \(x\), the same smoothed estimate \(\hat r(x)=\hat P(y=1\mid x)\) is used under all \(q\). Thus differences between \(F_m(\qemp)\), \(F_m(\qprod)\), and \(F_m(\qbal)\) reflect how the same conditional response is attributed under different state weightings, not changes in the response itself.

In simulations these weights include balanced weights and deliberately leakage-prone endpoints. In data these include empirical weights \(\qemp\), product weights \(\qprod\), and balanced weights \(\qbal\). The product weights are defined from the single-input rates in the local table,
\begin{equation}
  \qprod(x)=\prod_{i=1}^k p_i^{x_i}(1-p_i)^{1-x_i},
\end{equation}
where \(p_i=\sum_x q_{\rm emp}(x)x_i\). Balanced weights assign \(2^{-k}\) to each local state. For each state weighting \(q\), we primarily report the best order-\(m\) logistic model refit under that weighting. This tests whether the same conditional response table admits a first-order description under the specified state distribution. Separately, we report transfer analyses in which a model fit under \(\qemp\) is evaluated under \(\qbal\); these quantify distributional generalization of the empirical fit but are not used as the primary identifiability statistic. Available-entropy attribution is computed using Eq.~\ref{eq:Fm}.

For reweighted empirical tables, we also compute the effective sample size
\begin{equation}
  N_{\rm eff}
  =
  \frac{\left(\sum_t w_t\right)^2}{\sum_t w_t^2},
\end{equation}
where samples in state \(x\) receive weight proportional to \(q(x)/\qemp(x)\). In the CA1 three-input cubes, the median effective sample size for balanced weighting across folds is \(1{,}698\), with fifth and ninety-fifth percentiles \(790\) and \(3{,}346\), respectively.

\subsection{Local CA1 analysis}

The CA1 analysis uses the binarized hippocampal activity matrix made publicly available with \cite{Lynn2026}. For the local three-input analysis, target cells were selected across the activity-rate distribution, candidate inputs were chosen by positive output-input covariance on training bins, and all three-input subsets among the top candidates were tested. This intentionally focuses the diagnostic on the subset where both first-order predictability and higher-order-to-first leakage are expected to be strongest; it is not a random survey of all CA1 triplets. A subset was retained only if every one of its \(2^3\) input states had at least \(80\) samples in each training fold. Conditional response tables were estimated with Jeffreys smoothing,
\begin{equation}
  \hat r(x)
  =
  \frac{k_x+1/2}{n_x+1},
\end{equation}
where \(n_x\) is the number of samples in input state \(x\) and \(k_x\) is the number of output events in that state. The reported large-scale scan used \(40\) target cells, up to \(14\) candidate inputs per target, up to \(20\) retained subsets per target, and three cross-validation folds. This yielded \(776\) well-sampled local cubes.
We report these local analyses as diagnostics of identifiability, not as estimates of the fraction of all CA1 neurons with higher-order mechanisms.

\subsection{Matched additive-surrogate null}

To test whether the empirical-to-balanced reweighting split could be a finite-sample or smoothing artifact rather than genuine response structure, we constructed a matched additive null for each empirical-success cube. We fit the best additive (first-order) logistic response \(\hat r_{\rm add}(x)\) to the pooled observed table, then simulated output counts \(k_x^{\rm sim}\sim\mathrm{Binomial}(n_x,\hat r_{\rm add}(x))\) using the cube's own per-state counts \(n_x\) in each cross-validation fold. Each surrogate table was passed through the \emph{identical} pipeline -- the same Jeffreys smoothing, the same empirical/product/balanced reweighting, the same order-\(m\) refits, and the same \(F_1\ge 0.5\) classification -- and we recorded how often a genuinely additive table was classified as distribution-sensitive. We repeated this with \(20\) simulations per cube and report both the surrogate-table false-positive rate and the expected sensitive count under the null; the per-cube false-positive probabilities had median \(0\) and 95th percentile \(0\). The same additive surrogates calibrate the conditional log-odds contrast: we computed \(\max_{ij}|\Delta_{ij}|\) (median over folds) on each observed table and on \(200\) additive-surrogate tables per cube, and called a table conditionally non-additive when its observed value exceeded the per-cube null \(95\)th percentile. For the unselected base rate we removed the output-covariance selection: we drew input triplets that are well sampled by their own joint input states (well-sampledness depends only on the inputs, not the target) from among the most active neurons, paired each with randomly chosen targets, and ran the identical pipeline; we report statistics over the resulting empirical-success tables. The surrogate-null and unselected-base-rate scripts use fixed pseudorandom seeds (default seed \(0\)); the temporal leakage simulation uses fixed seeds generated from base seed \(7\) across the autocorrelation sweep.

\subsection{Visual-cortex comparison}

The visual-cortex natural-image and spontaneous recordings contain many neurons but fewer time bins than the CA1 recording, making well-sampled three-input cubes uncommon. The cross-dataset comparison therefore uses two-input local tables. The same reweighting logic applies: a pair interaction \(x_ix_j\) is the conditional response feature whose raw observable is the second-order coactivity \(\langle yx_ix_j\rangle\). To control for the larger visual population size, we repeated the visual analysis after downsampling V1 to \(N=1{,}485\) neurons and recomputed the CA1 comparison with the same two-input table procedure. This matched-population analysis controls neuron count but not recording duration; it is therefore a check on generality rather than a definitive CA1--V1 comparison.

\subsection{Raw coactivity prediction}

For a fitted first-order local model \(p_1(x)\), raw coactivities are evaluated as
\begin{equation}
  C_A(q,p_1)
  =
  \sum_x q(x)p_1(x)\prod_{a\in A}x_a.
\end{equation}
The corresponding measured value replaces \(p_1(x)\) by the empirical response table \(r(x)\). We report two versions of the balanced control. In the refit version, \(p_1(x)\) is the best first-order model fit under \(\qbal\). In the transfer version, \(p_1(x)\) is fit under \(\qemp\) and evaluated under \(\qbal\). The refit version asks whether raw coactivities are sufficient to diagnose conditional additivity; the transfer version asks whether the empirical first-order fit generalizes to a de-leaked input-state distribution.

Conditional interaction contrasts were computed from the same response tables. For pairwise contrasts, Eq.~\ref{eq:odds_contrast} was evaluated with probabilities clipped away from \(0\) and \(1\) for numerical stability. Unlike raw coactivities, these contrasts target additivity of the conditional log-odds.

\subsection{Temporal leakage simulation}

The temporal control used \(12\) binary inputs generated by independent Markov chains with stationary activity \(0.20\). To separate descriptive coactivity prediction from mechanism identification, we swept the lag-one input autocorrelation from approximately \(0.05\) to \(0.98\). The output response was generated from lagged inputs,
\begin{equation}
  P(y_t=1)
  =
  \sigmoid\left(b+\sum_i w_i x_i(t-1)\right),
\end{equation}
with no instantaneous drive. We fit two logistic models on the training portion of each simulated sequence: an instantaneous model using \(x_i(t)\) and a lagged model using \(x_i(t-1)\). On held-out time bins, we compared negative log likelihood and prediction of lagged coactivities \(\langle y_t x_i(t-1)\rangle\). This simulation is not intended as a full temporal model of any particular recording; it isolates the fact that autocorrelation can make instantaneous sufficient statistics predictive of delayed coactivities even when the ground truth is lagged.

\subsection{Feature leakage score}

For a higher-order input feature \(G(x)\) and first-order span \(\mathcal{F}_1=\operatorname{span}\{1,x_1,\ldots,x_k\}\), define
\begin{equation}
  R^2_{G\to 1}(q)
  =
  1-
  \frac{
    \min_{f\in\mathcal{F}_1}
    \E_q[(G(x)-f(x))^2]
  }{
    \Var_q(G)
  }.
\end{equation}
Large \(R^2_{G\to 1}\) means that the higher-order feature is largely predictable from first-order features under the input-state distribution. This is the finite-state projection version of the coskewness leakage calculation. We use ``leakage-prone'' descriptively for local tables in which this projection is large under empirical weights and small under product or balanced weights; in the selected CA1 cubes the median empirical value is \(0.380\), while product and balanced controls are near zero.

\section*{Acknowledgements}
This work has been made possible in part by a gift from the Chan Zuckerberg Initiative Foundation to establish the Kempner Institute for the Study of Natural and Artificial Intelligence at Harvard University.

\clearpage
\appendix
\section*{Supplementary Information}
\addcontentsline{toc}{section}{Supplementary Information}
\setcounter{figure}{0}
\renewcommand{\thefigure}{S\arabic{figure}}
\renewcommand{\theHfigure}{S\arabic{figure}}
\setcounter{table}{0}
\renewcommand{\thetable}{S\arabic{table}}
\renewcommand{\theHtable}{S\arabic{table}}

\section{Theory and identifiability results}

\begin{figure}[H]
  \centering
  \includegraphics[width=\textwidth]{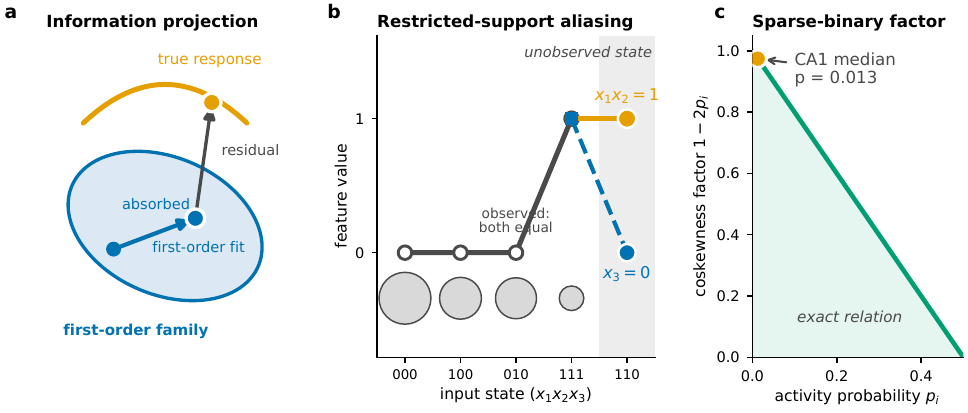}
  \caption{
  Geometric picture of the identifiability problem. (a) The restricted direct conditional MaxEnt model is an information projection onto the family of distributions generated by its first-order sufficient statistics. The true conditional response need not lie in this family. The component along an included sufficient statistic is \emph{absorbed} into the fitted direct parameters; only the component orthogonal to the included family (the \emph{residual}) is detected as remaining error. (b) On a restricted input support, an interaction feature \(x_ix_j\) and a single proxy input can take identical values at every sampled state. The two descriptions are then indistinguishable on observed data (gray circles below the axis have area proportional to the empirical \(P(x)\) of each input state; the unobserved state carries zero mass and so has no circle); the MaxEnt fit therefore cannot, by itself, identify the mechanism. (c) For sparse binary inputs, the projection of the centered product \(\delta_i\delta_j\) onto \(\delta_i\) is governed by the factor \(1-2p_i\) (Eq.~\ref{eq:binary_coskew}). At the median CA1 activity probability of \(0.013\), \(1-2p_i\approx 0.97\), converting ordinary input covariance almost directly into a coskewness channel that projects higher-order features onto first-order features.
  }
  \label{fig:supp_concept}
\end{figure}
\FloatBarrier

\subsection{Full derivation of the leakage formula}

Let the true conditional family be
\begin{equation}
  p_{\theta,\lambda}(y\mid x)
  =
  \frac{\exp\{\theta^\top F(y,x)+\lambda^\top G(y,x)\}}{Z_{\theta,\lambda}(x)}.
\end{equation}
The restricted fit \(\hat\theta(\lambda)\) solves the moment-matching equation
\begin{equation}
  \mu_F(\hat\theta(\lambda),0)
  =
  \mu_F(\theta_0,\lambda),
\end{equation}
where
\begin{equation}
  \mu_F(\theta,\lambda)
  =
  \E_{\Pemp(x)p_{\theta,\lambda}(y\mid x)}[F(y,x)].
\end{equation}
Differentiating with respect to \(\lambda\) at \(\lambda=0\) gives
\begin{equation}
  \frac{\partial\mu_F}{\partial\theta}
  \frac{d\hat\theta}{d\lambda}
  =
  \frac{\partial\mu_F}{\partial\lambda}.
\end{equation}
For a conditional exponential family,
\begin{align}
  \frac{\partial\mu_F}{\partial\theta}
  &=
  I_{FF}
  =
  \E_{\Pemp(x)}
  \left[
    \Cov_{p_{\theta_0}(y\mid x)}(F,F\mid x)
  \right],
  \\
  \frac{\partial\mu_F}{\partial\lambda}
  &=
  I_{FG}
  =
  \E_{\Pemp(x)}
  \left[
    \Cov_{p_{\theta_0}(y\mid x)}(F,G\mid x)
  \right].
\end{align}
Thus, when \(I_{FF}\) is invertible,
\begin{equation}
  \hat\theta-\theta_0
  =
  I_{FF}^{-1}I_{FG}\lambda
  +
  O(\|\lambda\|^2).
\end{equation}
This is the local omitted-statistic leakage formula used in the main text.

\subsection{Residual KL and the Schur complement}

Locally, the squared length of the omitted perturbation is
\begin{equation}
  \lambda^\top I_{GG}\lambda.
\end{equation}
The component projected into the included sufficient statistics is
\begin{equation}
  \lambda^\top I_{GF}I_{FF}^{-1}I_{FG}\lambda.
\end{equation}
The residual component orthogonal to the restricted family is therefore
\begin{equation}
  \lambda^\top I_{GG\mid F}\lambda
  =
  \lambda^\top
  \left(I_{GG}-I_{GF}I_{FF}^{-1}I_{FG}\right)
  \lambda.
\end{equation}
Consequently,
\begin{equation}
  \KL(p_{\theta_0,\lambda}\Vert p_{\hat\theta,0})
  =
  \frac12\lambda^\top I_{GG\mid F}\lambda
  +
  O(\|\lambda\|^3).
\end{equation}
This shows that the direct MaxEnt entropy can bound only the residual component of omitted structure, not the component absorbed into direct statistics.

\subsection{Moment constraints cannot rule out omitted statistics}

The following perturbation argument shows why matching the included moments is not enough to certify that omitted statistics are absent.

Let \(P_F^*(y\mid x)\) be the conditional MaxEnt solution constrained by the included statistics \(F(y,x)\) under an input-state distribution \(q(x)\). Consider a perturbation \(H(y,x)\) satisfying
\begin{equation}
  \E_{P_F^*(y\mid x)}[H(y,x)\mid x]=0
  \quad\text{for every }x,
\end{equation}
and
\begin{equation}
  \E_{q(x)P_F^*(y\mid x)}[F(y,x)H(y,x)]=0.
\end{equation}
For sufficiently small \(\epsilon\) such that \(1+\epsilon H(y,x)\ge 0\) on the observed support, define
\begin{equation}
  P_\epsilon(y\mid x)
  =
  P_F^*(y\mid x)\left[1+\epsilon H(y,x)\right].
\end{equation}
The first condition preserves conditional normalization and the second preserves all included MaxEnt constraints:
\begin{equation}
  \E_{q(x)P_\epsilon(y\mid x)}[F(y,x)]
  =
  \E_{q(x)P_F^*(y\mid x)}[F(y,x)].
\end{equation}
Therefore the restricted MaxEnt fit based on \(F\) is unchanged.

For an omitted statistic \(G(y,x)\),
\begin{equation}
  \E_{q(x)P_\epsilon(y\mid x)}[G(y,x)]
  =
  \E_{q(x)P_F^*(y\mid x)}[G(y,x)]
  +
  \epsilon
  \E_{q(x)P_F^*(y\mid x)}[G(y,x)H(y,x)].
\end{equation}
Whenever the last expectation is nonzero, the omitted statistic changes even though every included moment is identical. Thus constraints on \(F\) cannot, by themselves, establish that omitted statistics are absent.

\subsection{Count-valued direct statistics}

The same identifiability issue is not specific to binary variables. For count-valued outputs and inputs, a direct conditional MaxEnt model with full direct dependencies uses indicator-table sufficient statistics such as
\begin{equation}
  F_{i,a,r}(Y,X)=\mathbf{1}\{X_i=a,Y=r\}.
\end{equation}
Omitted pairwise input interactions use statistics such as
\begin{equation}
  G_{ij,a,b,r}(Y,X)=\mathbf{1}\{X_i=a,X_j=b,Y=r\}.
\end{equation}
The fitted direct model is an additive softmax model,
\begin{equation}
  P(Y=r\mid X=x)
  =
  \frac{\exp\{\alpha_r+\sum_i\phi_i(r,x_i)\}}
       {\sum_s\exp\{\alpha_s+\sum_i\phi_i(s,x_i)\}}.
\end{equation}
The same Fisher cross-covariance \(I_{FG}\) determines whether omitted count interactions project into direct count tables. Thus richer state alphabets change the sufficient statistics but do not remove the projection problem.

\section{Simulation parameters and scaling controls}

\begin{table}[H]
  \centering
  \caption{
  Ground-truth representation-sweep parameters. The generator has no first-order drive; the conditional log-odds are determined by centered pairwise products so that the direct model has no recoverable information under independent symmetric inputs.
  }
  \begin{tabular}{ll}
    \toprule
    Quantity & Value \\
    \midrule
    Number of inputs & \(K=32\) \\
    Interaction features & 120 randomly selected pairs \\
    Training samples & \(10^6\) \\
    Test samples & \(10^6\) \\
    Binary activity rate & 0.20 \\
    Target output rate & 0.15 \\
    Interaction strength & 3.0 \\
    Ground-truth features & centered pairwise products \\
    \bottomrule
  \end{tabular}
  \label{tab:synthetic_parameters}
\end{table}

\begin{table}[H]
  \centering
  \caption{
  Input representations used in the ground-truth sweep, with their theoretical skewness. All seven are deterministic (pointwise) transforms of a common latent Gaussian \(z_i\); most are monotone, while the half-normal (\(|z_i|\)) and chi-squared (\(z_i^2\)) maps are even. The lognormal scale is \(s=0.8\). For continuous positive representations, raw products contain mean-induced linear terms, so the interaction generator uses centered features when the goal is to test coskewness leakage rather than trivial mean leakage.
  }
  \begin{tabular}{lll}
    \toprule
    Representation & Transformation from latent \(z_i\) & Skewness \\
    \midrule
    Gaussian & \(x_i=z_i\) & \(0.00\) \\
    Uniform & \(x_i=\Phi(z_i)\) & \(0.00\) \\
    Half-normal & \(x_i=|z_i|\) & \(0.99\) \\
    Binary & \(x_i=\mathbf{1}\{z_i>\theta\}\) (rate 0.20) & \(1.50\) \\
    Exponential & \(x_i=-\ln[1-\Phi(z_i)]\) & \(2.00\) \\
    Chi-squared & \(x_i=z_i^2\) & \(2.83\) \\
    Lognormal & \(x_i=\exp(s z_i-s^2/2)\), \(s=0.8\) & \(3.69\) \\
    \bottomrule
  \end{tabular}
  \label{tab:input_representations}
\end{table}

\begin{table}[H]
  \centering
  \caption{
  Fraction of recoverable information captured by the direct model in the representation sweep, for all seven representations plotted in Fig.~\ref{fig:main_controls}a. The same higher-order generator can yield almost no direct attribution (symmetric Gaussian/uniform), intermediate attribution (skewed positive representations), or high direct attribution (binary), depending only on representation and correlation.
  }
  \begin{tabular}{cccccccc}
    \toprule
    \(\rho\) & Binary & Gaussian & Uniform & Lognormal & Half-normal & Exponential & Chi-squared \\
    \midrule
    0.0 & 0.04 & 0.00 & 0.00 & 0.07 & 0.02 & 0.05 & 0.07 \\
    0.1 & 0.11 & 0.00 & 0.00 & 0.11 & 0.02 & 0.08 & 0.07 \\
    0.2 & 0.18 & 0.00 & 0.00 & 0.15 & 0.04 & 0.13 & 0.10 \\
    0.3 & 0.25 & 0.00 & 0.00 & 0.19 & 0.08 & 0.18 & 0.13 \\
    0.5 & 0.42 & 0.00 & 0.00 & 0.26 & 0.18 & 0.28 & 0.22 \\
    0.7 & 0.61 & 0.00 & 0.00 & 0.33 & 0.32 & 0.39 & 0.37 \\
    0.9 & 0.84 & 0.00 & 0.00 & 0.42 & 0.45 & 0.56 & 0.65 \\
    \bottomrule
  \end{tabular}
  \label{tab:phi_direct_sweep}
\end{table}

\begin{figure}[H]
  \centering
  \includegraphics[width=\textwidth]{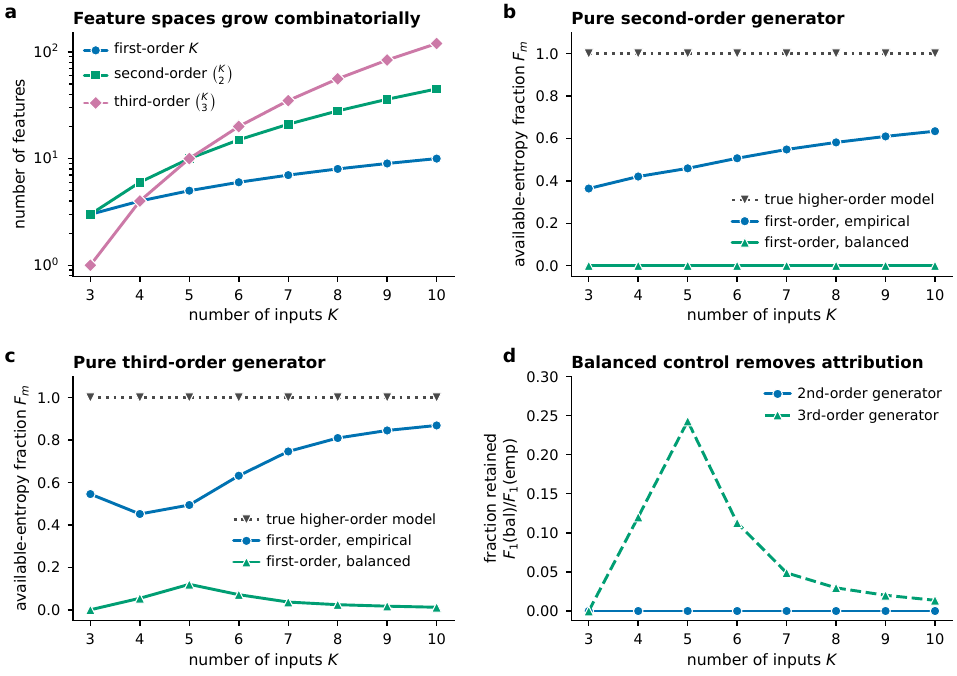}
  \caption{
  Scaling of first-order attribution with input count under a leakage-prone empirical distribution. \textbf{a} The number of \(m\)th-order features over \(K\) binary inputs grows combinatorially, so a larger population provides more possible proxies for omitted higher-order structure. \textbf{b,c} For pure second- and third-order generators with no first-order drive, empirical first-order entropy attribution rises with \(K\), while balanced first-order attribution stays at or near zero and the correctly specified higher-order model recovers the response. \textbf{d} The fraction of empirical first-order attribution retained after balanced reweighting---the ratio \(F_1(\qbal)/F_1(\qemp)\), shown for the second- and third-order generators---stays modest (peaking near \(0.25\) for the third-order generator around \(K=5\)) and does not grow with \(K\). Adding inputs therefore does not automatically restore the first-order interpretation.
  }
  \label{fig:supp_scaling}
\end{figure}
\FloatBarrier

\section{Additional empirical controls and diagnostics}

\begin{figure}[H]
  \centering
  \includegraphics[width=0.82\textwidth]{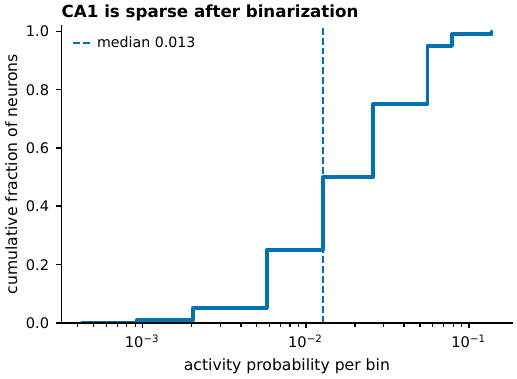}
  \caption{
  CA1 activity sparsity. Cumulative distribution of single-neuron activity probabilities (fraction of time bins active) in the binarized CA1 recording; the population is sparse, with median \(0.013\) and mean \(0.019\). At this sparsity the binary coskewness factor \(1-2p_i\approx 0.97\) (Eq.~\ref{eq:binary_coskew}; Fig.~\ref{fig:supp_concept}c), so ordinary input covariance is converted almost directly into a coskewness channel. The coskewness-projection enrichment of the selected pairs, the removal of higher-order feature leakage under product and balanced weights, and the response-dependence of the entropy drop are shown in the main text (Fig.~\ref{fig:ca1_entropy}a--c) and are not duplicated here.
  }
  \label{fig:supp_ca1_coskew}
\end{figure}

\begin{figure}[H]
  \centering
  \includegraphics[width=\textwidth]{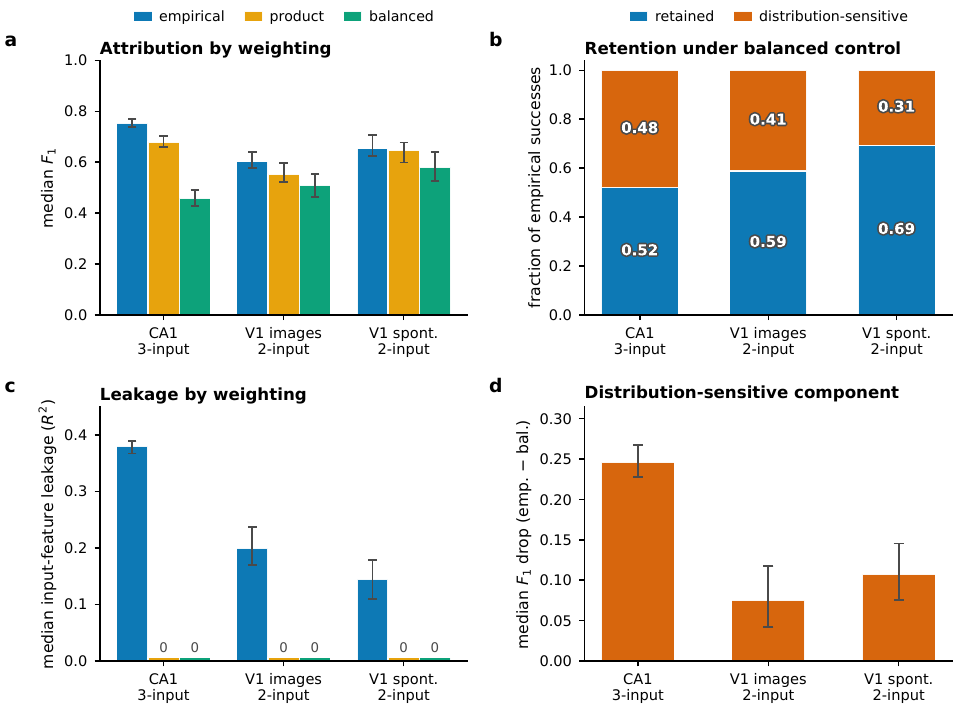}
  \caption{
  Cross-dataset local reweighting: CA1 (three-input cubes) versus V1 natural-image and V1 spontaneous activity (two-input tables). (a) Median local \(F_1\) under empirical, product and balanced weights. (b) Fraction of empirical successes retained versus distribution-sensitive under balanced control. (c) Median input-feature leakage under each weighting. (d) Median \(F_1\) drop (empirical \(-\) balanced), the distribution-sensitive component, largest in CA1. All datasets retain a measurable distribution-sensitive component.
  }
  \label{fig:supp_cross_dataset}
\end{figure}

\begin{figure}[H]
  \centering
  \includegraphics[width=\textwidth]{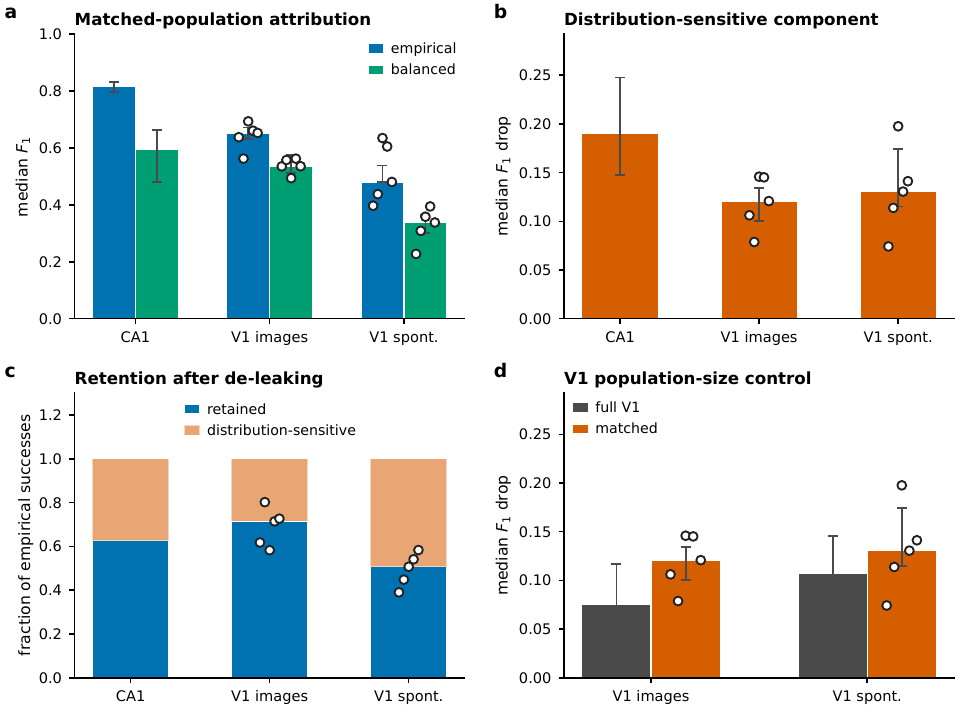}
  \caption{
  Matched-population control, downsampling V1 to the CA1 size of \(N=1{,}485\) neurons with two-input local tables for both datasets. (a) Matched-population median \(F_1\) (empirical versus balanced); dots are per-seed values. (b) Distribution-sensitive component under matching. (c) Retention after de-leaking. (d) V1 population-size control comparing full V1 with the \(N=1{,}485\) matched subsample. The cross-dataset comparison is not explained only by V1 population size.
  }
  \label{fig:supp_matched_population}
\end{figure}

These controls support the empirical analyses without changing the main diagnostic: the local response table is held fixed while the state weights are changed. Figure~\ref{fig:supp_ca1_coskew} documents the CA1 activity sparsity that places the recording in the sparse-binary regime; Fig.~\ref{fig:supp_cross_dataset} compares CA1 with visual-cortex natural-image and spontaneous recordings; and Fig.~\ref{fig:supp_matched_population} repeats the comparison after downsampling V1 to the CA1 population size (\(N=1{,}485\) neurons).

\FloatBarrier

\FloatBarrier

\FloatBarrier

\begin{figure}[H]
  \centering
  \includegraphics[width=\textwidth]{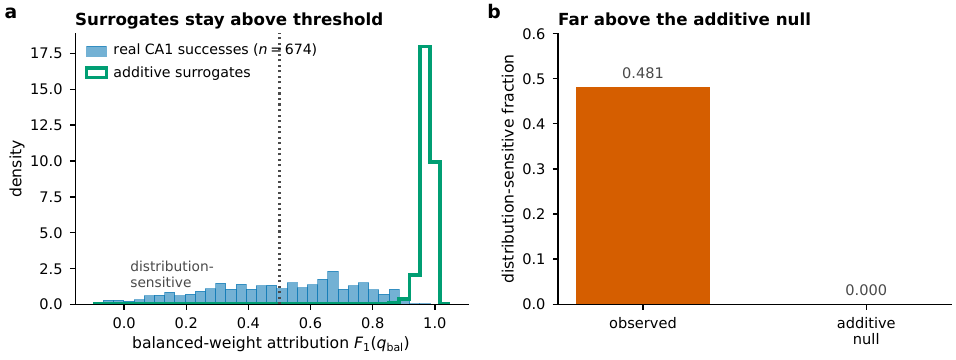}
  \caption{
  Matched additive-surrogate null for the CA1 distribution-sensitive split. For each empirical-success cube we fit the best additive (first-order) response, simulated spike counts at the observed per-state counts and cross-validation folds, and pushed each surrogate through the identical Jeffreys-smoothing, reweighting and classification pipeline. (a) Distribution of balanced-weight attribution \(F_1(\qbal)\) for the real empirical-success cubes (blue), which has a tail below the \(F_1=0.5\) success threshold (dotted line)---the distribution-sensitive tables---overlaid on the additive-surrogate distribution (green), which stays above threshold. (b) The observed distribution-sensitive fraction versus the additive-surrogate null: genuinely additive tables are essentially never misclassified, so the observed split is not a finite-sample or smoothing artifact.
  }
  \label{fig:supp_surrogate}
\end{figure}
\FloatBarrier

\section{Temporal leakage schematic}

\begin{figure}[H]
  \centering
  \includegraphics[width=\textwidth]{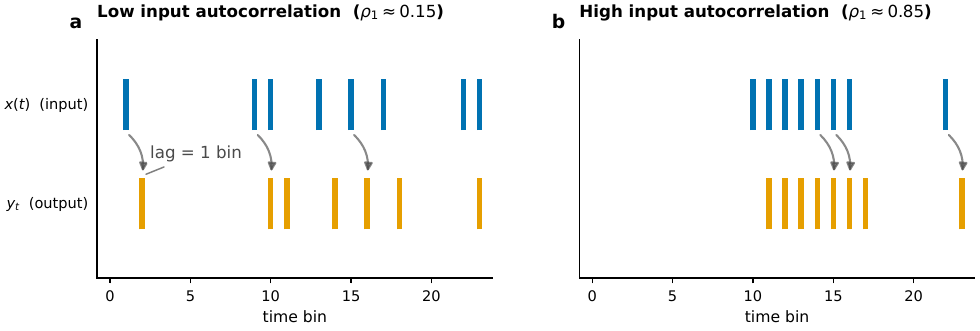}
  \caption{
  Temporal leakage from autocorrelated inputs. In this schematic the ground-truth rule is \(y_t=x(t-1)\), so each output spike (orange) is the input (blue) shifted by one bin (gray arrows). \textbf{a} At low input autocorrelation, \(x(t)\) is a poor proxy for \(x(t-1)\), and an instantaneous fit \(\hat y_t=f(x(t))\) differs clearly from the correctly lagged fit. \textbf{b} At high autocorrelation, \(x(t)\approx x(t-1)\), so an instantaneous fit can predict delayed coactivities even though the response rule is lagged. The corresponding quantitative control---instantaneous delayed-coactivity \(R^2\) together with the held-out NLL gain of the lagged model---is shown in Fig.~\ref{fig:nature_fig4_control}g,h.
  }
  \label{fig:supp_visual_temporal}
\end{figure}

Temporal correlations in the inputs can make an instantaneous model pass a delayed-coactivity test even when the response is lagged. Figure~\ref{fig:supp_visual_temporal} shows the minimal case: the output depends on the previous input bin, but high input autocorrelation makes the current bin a close proxy for the delayed one. The schematic is meant only as a visual guide; the quantitative temporal leakage analysis is shown in Fig.~\ref{fig:nature_fig4_control}g,h.

\FloatBarrier

\end{document}